\documentclass[twocolumn,pre,superscriptaddress,showpacs,byrevtex]{revtex4}
\usepackage{epsf,graphicx,amssymb}
\begin{document}
\draft
\title{Simulations of vibrated granular medium\\ with impact velocity dependent  restitution coefficient}
\author{Sean McNamara}
\affiliation{Centre Europ\'een de Calcul Atomique et Mol\'eculaire, 46, all\'ee d'Italie, 69 007 Lyon, France}
\affiliation{Permanent address: I.C.P., Universit\"at Stuttgart, 70569 Stuttgart, Germany}
\author{Eric Falcon}
  \email[Email address: ]{Eric.Falcon@ens-lyon.fr}
  \homepage{http://perso.ens-lyon.fr/eric.falcon/}
\affiliation{Laboratoire de Physique, \'Ecole Normale Sup\'erieure de Lyon, UMR 5672, 46, all\'ee d'Italie, 69 007 Lyon, France}
\date{\today}

\begin{abstract}
We report numerical simulations of strongly vibrated granular materials designed to mimic recent experiments performed both in presence or absence of gravity.  The coefficient of restitution used here depends on the impact velocity by taking into account both the viscoelastic and plastic deformations of particles, occurring at low and high velocities respectively. We show that this model with impact velocity dependent restitution coefficient reproduce results that agree with experiments. We measure the scaling exponents of the granular temperature, collision frequency, impulse, and pressure with the vibrating piston velocity as the particle number increases. As the system changes from a homogeneous gas state at low density to a clustered state at high density, these exponents are all found to decrease continuously with the particle number. All these results differ significantly from classical inelastic hard sphere kinetic theory and previous simulations, both based on a constant restitution coefficient. 
\end{abstract}
\maketitle

\section{Introduction}
The past decade has seen the publication of many experimental \cite{Warr:95,Yang:00,Luding:94a}, numerical \cite{Luding:94a,Luding:94b,McNamara:98}, and theoretical \cite{Warr:95,Huntley:98,Kumaran:98,Lee:95} studies of strongly vibrated granular media. This problem is interesting because vibrated granular media
are simple but nontrivial example of a nonequilibrium steady states and the only way to experimentally realize granular gases \cite{GranularGases}. However, numerous questions remain about the link between experiments on one hand, and theory and simulations on the other.  Most numerical and theoretical studies were not intended to be compared with experiments.  Therefore, they have parameter values far from the experimental ones, and none of them predict even the most basic features of the experimental results.  

In this paper, we bridge the gap between experiments and numerics by presenting simulations of strongly vibrated granular materials designed to mimic recent experiments performed both in presence \cite{Falcon:01} or absence \cite{Falcon:99} of gravity. We present the first simulations which resemble the experiments for a large range of parameters. We show that two parameters are especially important for the agreement between experiment and simulation. First of all, the coefficient of restitution has to be dependent on the particle impact velocity by taking into account both the viscoelastic and plastic deformations of particles occurring at low and high velocities respectively. Most previous numerical studies consider only constant restitution coefficient \cite{Luding:94a,Luding:94b,McNamara:98}, or few studies with the slightly velocity dependence (due the only viscoelastic contribution) \cite{Brilliantov:00}. Secondly, it is important to explicitly consider the number of particles $N$.  Studying only one value of $N$ or comparing results obtained at different $N$ can lead to interpretive difficulties. 

Beyond these agreements between experiments and our simulations, we find new results that differ significantly from classical inelastic hard sphere kinetic theory and previous simulations. We measure the scaling exponents of the granular temperature, collision frequency, impulse, and pressure with the vibrating piston velocity as the particle number increases, both in the presence or absence of gravity. We show that the system undergoes a smooth transition from a homogeneous gas state at low density to a clustered state at high density.

The paper has the following structure. In Sec.~\ref{descrp}, we present a description of the simulations (notably the model of impact velocity dependent restitution coefficient, and the influence of other simulations parameters). Section~\ref{McNFal:sec:experiment} provides a comparison of simulations and experiments (showing the importance of the variable coefficient of restitution and the particle number), and the results of the scaling exponents. Section \ref{otherparameters} focus on the influence of other simulations parameters (bed height, box size, particle rotations, gravity). Finally, in Sec.~\ref{conclusions} we summarize our results.  

\section{Description of the simulations}
\label{descrp}
\subsection{The variable coefficient of restitution}
The greatest difference between our simulations and the previous numerical studies of vibrated granular media \cite{Luding:94a,Luding:94b,McNamara:98} is that we use a restitution coefficient that depends on impact velocity.  The restitution coefficient $r$ is the ratio between the relative normal velocities before and after impact.  In all previous simulations of strongly vibrated granular media, the coefficient of restitution is considered to be constant and lower than 1.  However, since a century, it has been shown from impact experiments that $r$ is a function of the impact velocity $v$ \cite{Raman:1918,Johnson:85,Labous:97,Kuwabara:87,Falcon:98}. Indeed, for
metallic particles, when $v$ is large ($v \gtrsim 5$ m/s \cite{Johnson:85}), the colliding particles deform fully plastically and $r \propto v^{-1/4}$ \cite{Raman:1918,Johnson:85,Labous:97}. When $v\lesssim 0.1$ m/s \cite{Johnson:85}, the deformations are elastic with mainly viscoelastic dissipation, and $1-r \propto v^{1/5}$ \cite{Labous:97,Kuwabara:87,Falcon:98,Hertzsch:95}. Such velocity-dependent restitution coefficient models have recently shown to be important in numerical \cite{Brilliantov:00,Saluena:99,Bizon:98,Poschel:01,Goldman:98,Salo:88} and experimental \cite{Falcon:98,Bridges:84} studies. Applications include: granular fluidlike properties (convection \cite{Saluena:99}, surface waves \cite{Bizon:98}), collective collisional processes (energy transmission \cite{Poschel:01}, absence of collapse \cite{Falcon:98,Goldman:98}) and planetary rings \cite{Salo:88,Bridges:84}.  But surprisingly, such model has not yet been tested numerically for strongly vibrated granular media.

\begin{figure}[htb]
\centerline{
  \epsfysize=60mm \epsffile{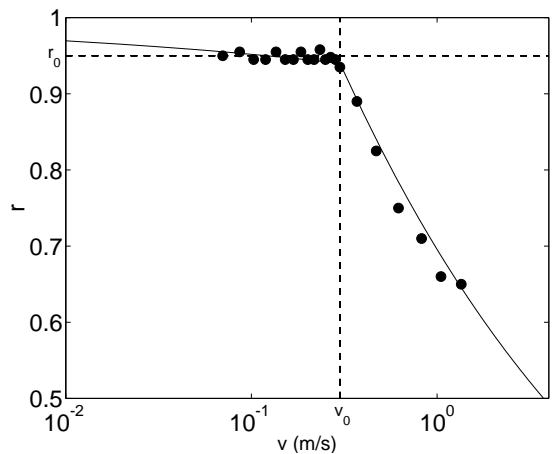}
}
\caption[]{The restitution coefficient $r$ as
 a function
of impact velocity $v$, as given in Eq.~(\ref{McNFal:eq:rvimp}) (solid line).  The dashed lines
 show $v_0=0.3$ m/s and $r_0=0.95$. Experimental points ($\bullet$) for
steel spheres were extracted from Fig.1 of Ref. \cite{Lifshitz:64}}
\label{McNFal:fig:rvimp}
\end{figure}

In this paper, we use a velocity dependent restitution coefficient  $r(v)$ and join the two regimes of dissipation (viscoelastic and plastic) together as simply as possible, assuming that
\begin{equation}
r(v) = \left\{ \matrix{
1-(1-r_0) \left(\frac{v}{v_0} \right)^{1/5}, &  v \le v_0, \cr 
r_0 \left( \frac{v}{v_0} \right)^{-1/4}, & v \ge v_0 ,}\right.
\label{McNFal:eq:rvimp}
\end{equation}
where $v_0 = 0.3$ m/s is chosen, throughout the paper, to be the yielding velocity for stainless steel particles \cite{Johnson:85,Lifshitz:64} for which $r_0$ is close to 0.95 \cite{Lifshitz:64}. Note that $v_0 \sim 1/\sqrt{\rho}$ where $\rho$ is the density of the sphere \cite{Johnson:85}. We display in Fig.~\ref{McNFal:fig:rvimp} the velocity dependent restitution coefficient of Eq.~(\ref{McNFal:eq:rvimp}), with $r_0=0.95$ and $v_0 = 0.3$ m/s, that agrees well with experimental results on steel spheres from Ref.\ \cite{Lifshitz:64}. As also already noted by Ref.\ \cite{Johnson:85}, the impact velocity to cause yield in metal surfaces is indeed relatively small.  For metal, it mainly comes from the low yield stress value ($Y \sim 10^9$ N/m$^2$) with repect to the elastic the Young modulus ($E \sim 10^{11}$ N/m$^2$). Most impacts between metallic bodies thus involve some plastic deformation. 

\subsection{The other simulation parameters}

The numerical simulation consists of an ensemble of identical hard disks
of mass $m \approx 3 \times 10^{-5}$ kg
excited vertically by a piston in a two-dimensional box. Simulations are
done both in the presence ($g=9.8$ m/s$^2$) and absence ($g=0$) of uniform gravity $g$. Collisions
are assumed instantaneous and thus only binary collisions occur. For
simplicity, we neglect the rotational degree of freedom.  
Collisions with the wall are treated in the same way as collisions between
particles, except the wall has infinite mass. 

Motivated by recent 3D experiments on stainless steel spheres, 2 mm in
diameter, fluidized by a vibrating piston \cite{Falcon:01}, we choose the
simulation parameters to match the experimental ones: in the simulations, the vibrated piston
at the bottom of the box has amplitude $A=25$ mm (distance between the highest and lowest positions of the piston) and frequencies
$5\;\text{Hz} \le f \le 50\;\text{Hz}$.  The piston is nearly sinusoidally vibrated with
a waveform made by joining two parabolas together. The vertical displacement of the piston $z(t)$ during time $t$ then is $z(t)=\frac{A}{2}(t^2-t_0^2)$ for $-t_0\leq t \leq t_0$ and $z(t)=-\frac{A}{2}(t^2-t_0^2)$ for $t_0\leq t \leq 3t_0$ with $t_o=1/(4f)$. This leads to a maximum piston velocity given by $V=4Af$.
The particles are disks 
$d = 2\;\text{mm}$
in diameter with stainless steel collision properties through $v_0$ and
$r_0$ (see Fig.~\ref{McNFal:fig:rvimp}). 
The box
has width $L=20$ cm and horizontal periodic boundary conditions. 
Since our simulations are two
dimensional,  we consider the simulation geometrically equivalent to the
experiment when their number of layers of particles, $n=Nd/L$, are equal. 
Hence in the simulation, a layer of particles, $n=1$, corresponds to $100$
particles.  We checked that $n$ is an appropriate way to measure
the number of particles by also running simulations at $L=10$ cm
and $L=40$ cm.  None of this paper's results depend significantly
on $L$.
As in the experiments, the height $h$ of the box depends on the
number of particles in order to have a constant difference $h-h_0=15$ mm,
where $h_0$ is the height of the bed of particles at rest. Heights are
defined from the piston at its highest position. The influence of $h-h_0$ on the results is discussed in Sec.~\ref{otherparameters}.

\begin{figure}[htb]
\centerline{
\begin{tabular}{cc}
{\bf (a) experiments} & {\bf (b) $r=r(v)$}\\
 \includegraphics[width=.225\textwidth]{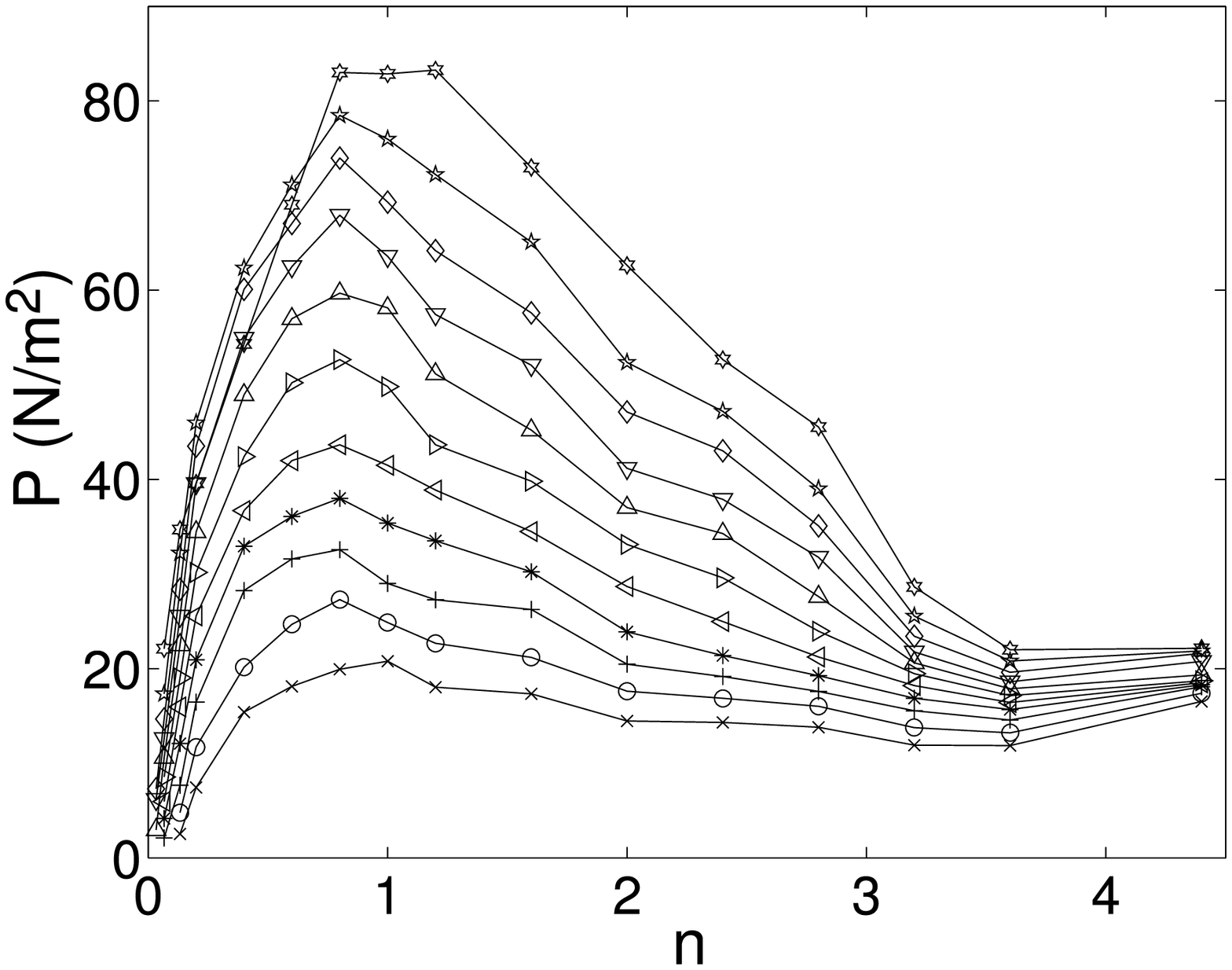} & \includegraphics[width=.225\textwidth]{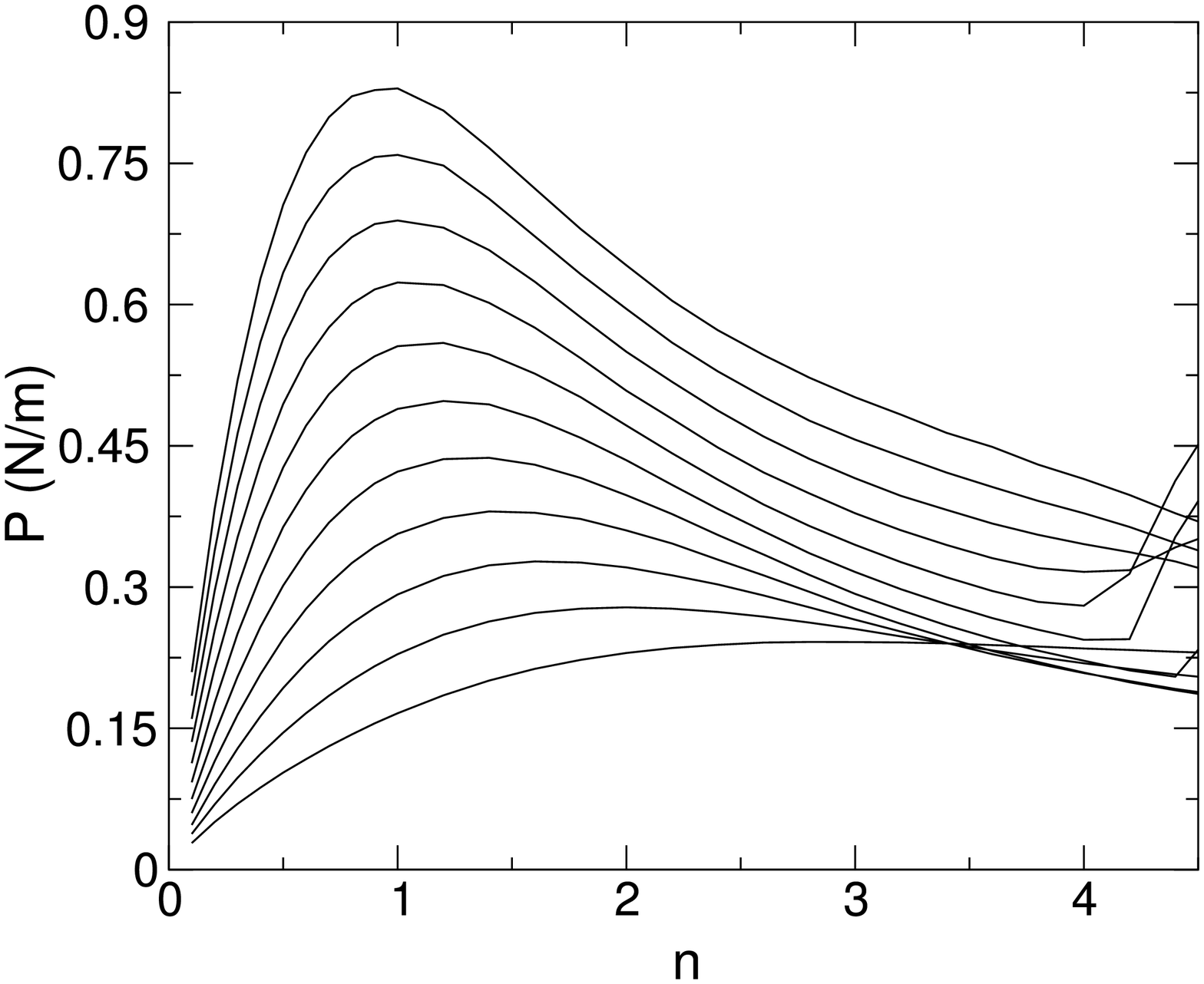}\\
{\bf (c) $r=0.95$} & {\bf (d) $r=0.7$}\\
\includegraphics[width=.225\textwidth]{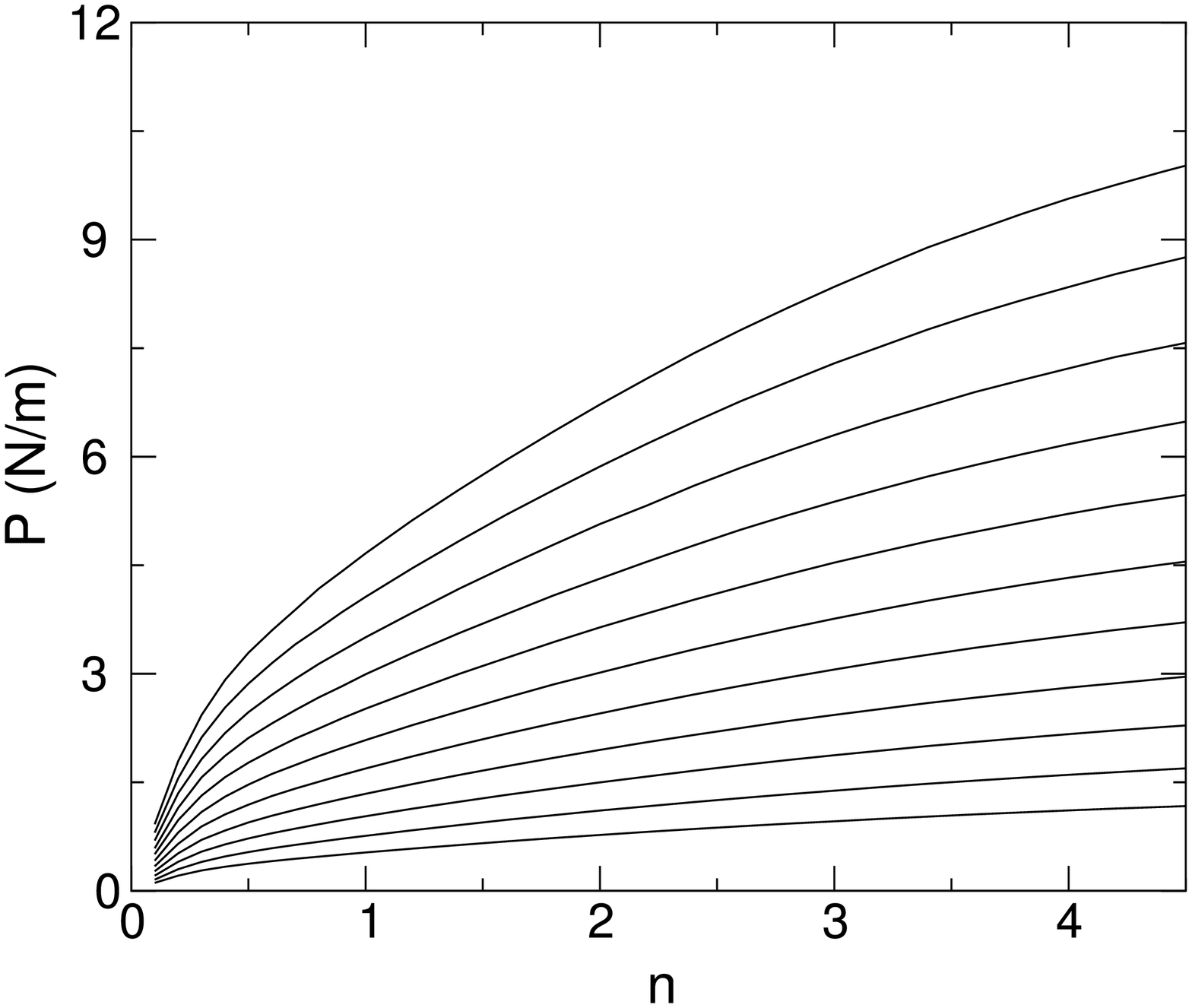}& \includegraphics[width=.225\textwidth]{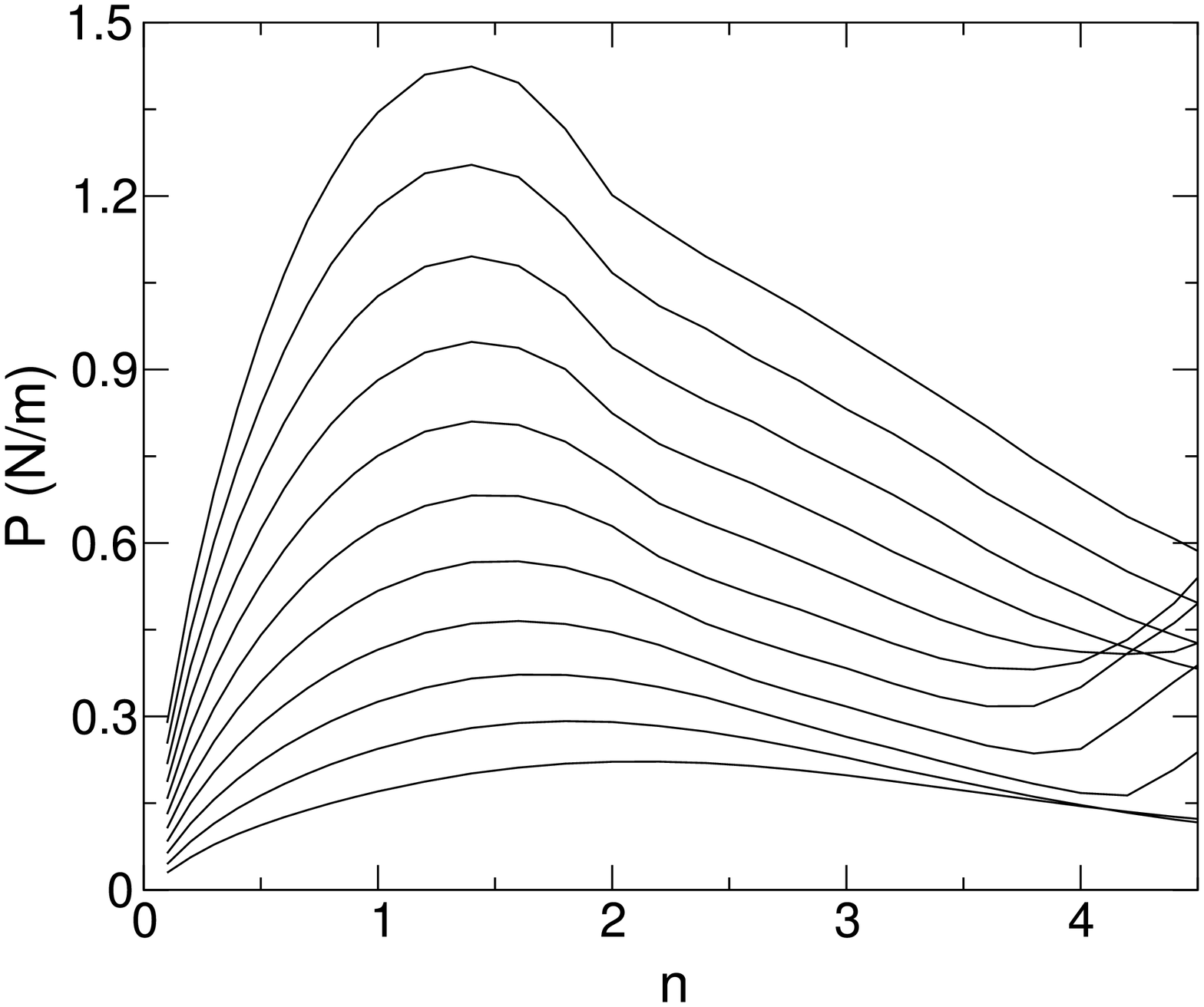}\\
\end{tabular}
}
\caption[]{Time averaged pressure $P$ on the top of the cell 
as a function of particle layer, $n$, for various vibration frequencies,
$f$ : {\bf (a)} Experimental results from \cite{Falcon:01} for stainless
steel beads 2 mm in diameter, with $A=25$ mm, $10\le f \le 20 Hz$ with a 1
Hz step (from lower to upper)  and $h-h_0 = 5$ mm. {\bf (b)} Numerical
simulation where the coefficient of restitution is given by
Eq.~(\ref{McNFal:eq:rvimp}), {\bf (c)} Numerical simulation with a coefficient of
restitution is $0.95$, independent of impact velocity,
{\bf (d)} Numerical simulation with a coefficient of
restitution is $0.7$, independent of impact velocity. The simulations
(b) -- (d) are 2D with gravity, done for 2 mm disks,
with $A=25$ mm, $10\le f \le 30$ Hz with a 2 Hz step
(from lower to upper) and $h-h_0 = 15$ mm.  In the simulations, the
two-dimensional pressure is given in N/m.}
\label{McNFal:fig:freqscan}
\end{figure}


\section{Comparison of simulation and experiment - Scaling properties}
\label{McNFal:sec:experiment}

\subsection{The importance of the variable coefficient of restitution}

We examine first the dependence of the pressure on
the number of particle layers for maximum velocity of the piston
$1 \lesssim V\lesssim 5$ m/s ($V=4Af$). The time averaged pressure at
the upper wall is displayed in Fig.~\ref{McNFal:fig:freqscan} as a function of $n$
for various $f$: from the experiments of Falcon et al. \cite{Falcon:01} (see
Fig.~\ref{McNFal:fig:freqscan}a), from our simulations with velocity dependent
restitution coefficient $r=r(v)$ proposed in Eq.~(\ref{McNFal:eq:rvimp}) (see
Fig.~\ref{McNFal:fig:freqscan}b), and with constant restitution coefficient
$r=0.95$, often used to describe steel particles (see
Fig.~\ref{McNFal:fig:freqscan}c), and finally with an unrealistic constant restitution coefficient
$r=0.7$ (see Fig.~\ref{McNFal:fig:freqscan}d).  
Simulations with $r=r(v)$ give results in
agreement with the experiments: At constant external driving, the pressure
in both Figs.~\ref{McNFal:fig:freqscan}a and \ref{McNFal:fig:freqscan}b passes through a
maximum for a critical value of $n$ roughly corresponding to one particle
layer.  For $n<1$, most particles are in vertical ballistic motion between
the piston and the lid. Thus, the mean pressure increases roughly
proportionally to $n$. When $n$ is increased such that $n>1$, interparticle
collisions become more frequent. The energy dissipation is increased and
thus the pressure decreases.  This maximum pressure is not due to
gravity because it also appears in simulations with $g=0$ and $r=r(v)$. 
Furthermore, the maximum persists when $g$ is increased above $9.8$ m/s$^2$.  For
$n \ge 4$ and for certain frequencies, a resonance appears in
Fig.~\ref{McNFal:fig:freqscan}b which is controlled by the ratio between the
vibration period and the particle flight time under gravity,
$\sqrt{g/h}/f$. Turning our attention to Fig.~\ref{McNFal:fig:freqscan}c, we see
that setting $r=0.95$ independently of impact velocity gives pressure
qualitatively different from experiments. The difference between
Figs.~\ref{McNFal:fig:freqscan}b and \ref{McNFal:fig:freqscan}c can be understood by
considering a high velocity collision (e.g. $v=1$ m/s). In
Fig.~\ref{McNFal:fig:freqscan}b, this collision has a restitution coefficient of
$r=r(1$ m/s$) \approx 0.7$ (see Fig.~\ref{McNFal:fig:rvimp}), whereas in
Fig.~\ref{McNFal:fig:freqscan}c, $r$ is fixed at 0.95 for all collisions. This
means that for equal collision frequencies, dissipation is much stronger
for $r=r(v)$ than for $r=0.95$, because the high velocity collisions
dominate the dissipation. Stronger dissipation leads to lower granular
temperatures and thus to lower pressures. 

We can check this interpretation
by changing the constant restitution coefficient to $r=0.7$ and then
comparing it to $r=r(v)$. In these two cases, the high velocity collisions
will have roughly the same restitution coefficient. We indeed observed a
pressure that decreases for large $n$ for constant $r=0.7$
(see Fig.~\ref{McNFal:fig:freqscan}d). 
Therefore, surprisingly, constant $r=0.7$ reproduces more precisely the experimental
pressure measurements than constant $r=0.95$, even though $r=0.95$ or
$r=0.9$ is often given as the restitution coefficient of steel. However, if we look at other properties, we see that $r=0.7$ and $r=r(v)$ give very different predictions.  
\begin{figure}[ht]
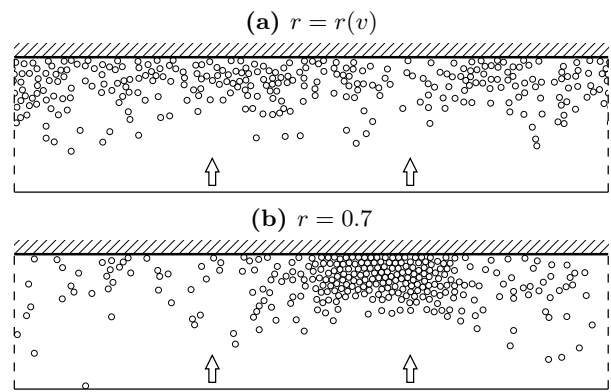

\centerline{
\begin{tabular}{c}
{\bf (a) $r=r(v)$} \\
\includegraphics[width=0.45\textwidth]{fig03a.eps}\\
{\bf (b) $r=0.7$} \\
\includegraphics[width=0.45\textwidth]{fig03b.eps}\\
\end{tabular}
}
\caption[]{Snapshots from the simulations with $n=3$, gravity $g\ne0$, driving frequency $f=30\;\text{Hz}$ and $h-h_0=15\;\text{mm}$.
The upper wall is stationary, and the lower wall is the piston,
which is at its lowest position in both snapshots.  The horizontal
boundaries are periodic (indicated by dashed lines).  Gravity points
downwards.  {\bf (a)} $r=r(v)$, as given in Eq.~(\ref{McNFal:eq:rvimp}),
and {\bf (b)} constant $r=0.7$.  
In (b) we see a tight cluster which was not observed in the experiments.}
\label{McNFal:fig:snapshots}
\end{figure}

For example, in Fig.~\ref{McNFal:fig:snapshots}, we show two snapshots from two different simulations one with $r=r(v)$ and another with $r=0.7$, both with $n=3$ in the presence of gravity.  When $r=r(v)$, the particles are concentrated in the upper half of the chamber, but they are evenly spread in the horizontal direction (see Fig.~\ref{McNFal:fig:snapshots}a). The system is hotter and less dense near the vibrating wall, and colder and denser by the opposite wall. But, when $r=0.7$, the majority of the particles are confined to a tight cluster, pressed against the upper wall, coexisting with low density region (see Fig.~\ref{McNFal:fig:snapshots}c).  This instability has been already
been reported numerically \cite{Argentina:02}, although for much different
parameters (constant restitution coefficient $r=0.96$, thermal walls, 
no gravity, and large $n$). However, nothing like this was ever seen experimentally. Therefore, if one is seeking information about particle positions, $r=0.7$ gives incorrect
results even though it gives acceptable results for the pressure. We conclude, therefore, that the only way to successfully describe all the properties
in all situations is to use a velocity dependent restitution
coefficient model.  

\subsection{The importance of the particle number}

\begin{figure*}[ht]
\centerline{
\begin{tabular}{cc}
{\bf (a) $r=0.95$, $g=0$} & {\bf (b) $r=r(v)$, $g=0$}\\
 \epsfysize=46mm \epsffile{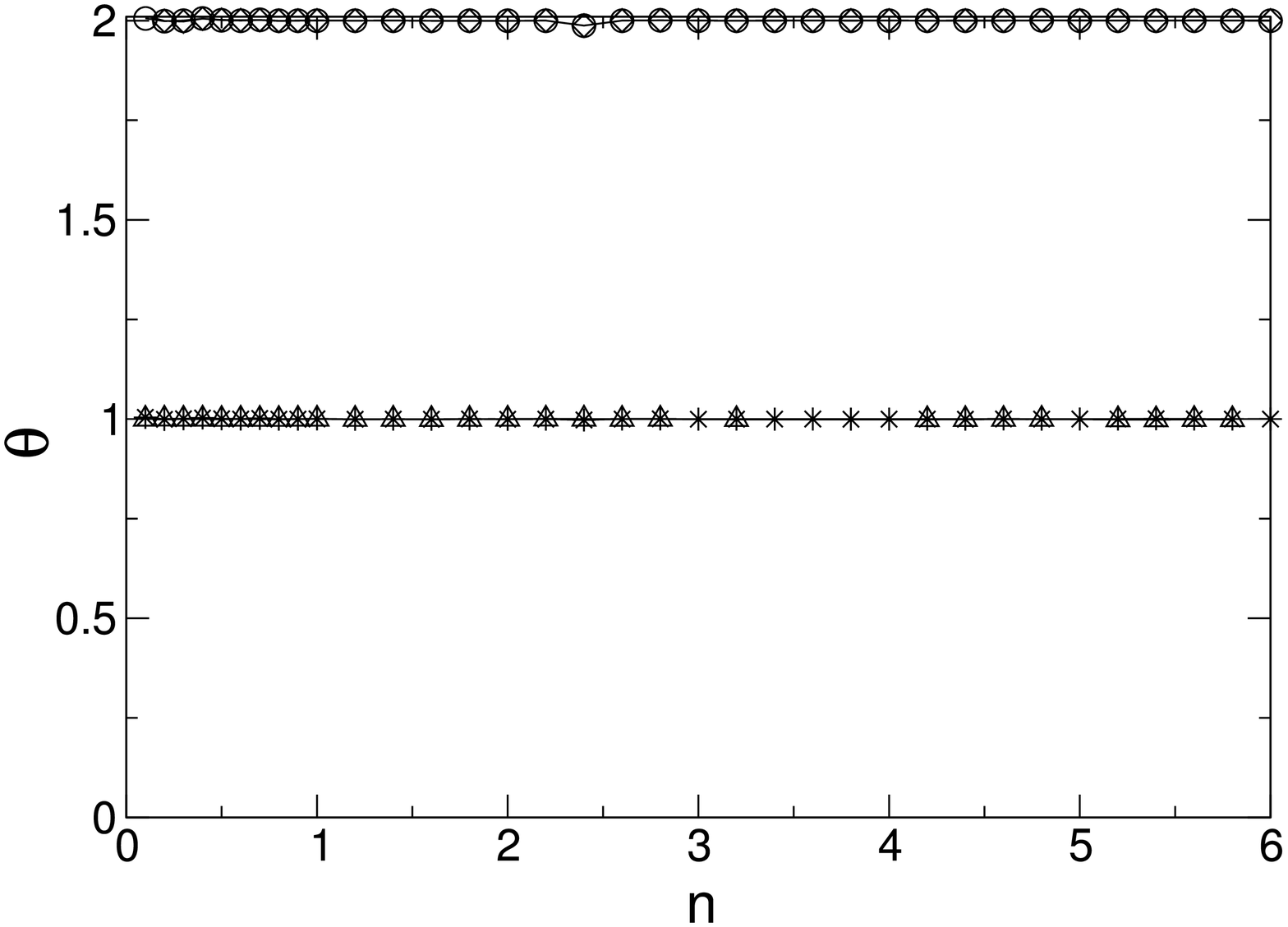} &
 \epsfysize=46mm \epsffile{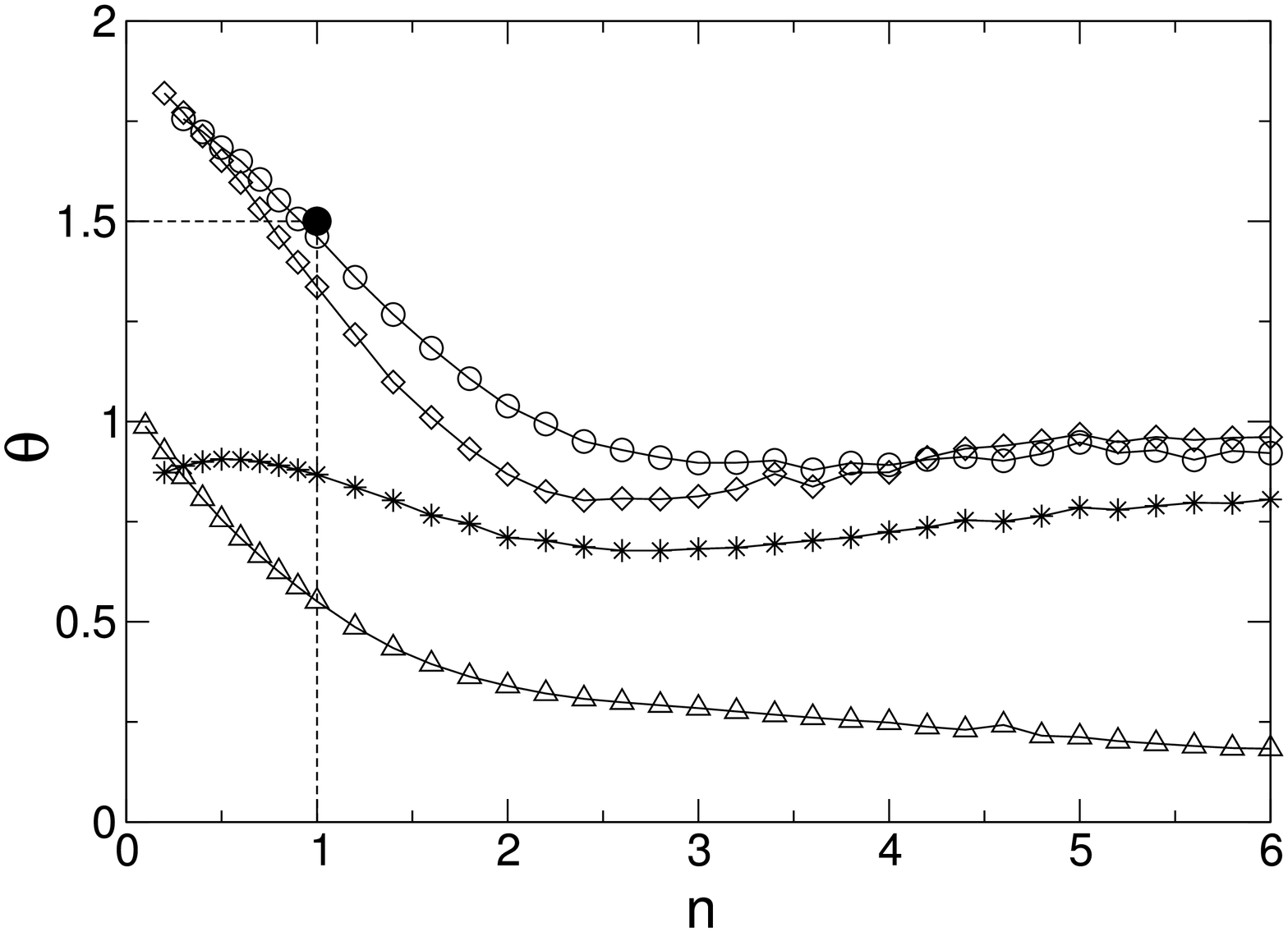} \\
{\bf (c) $r=0.95$, $g$ } & {\bf (d) $r=r(v)$, $g$}\\
 \epsfysize=46mm \epsffile{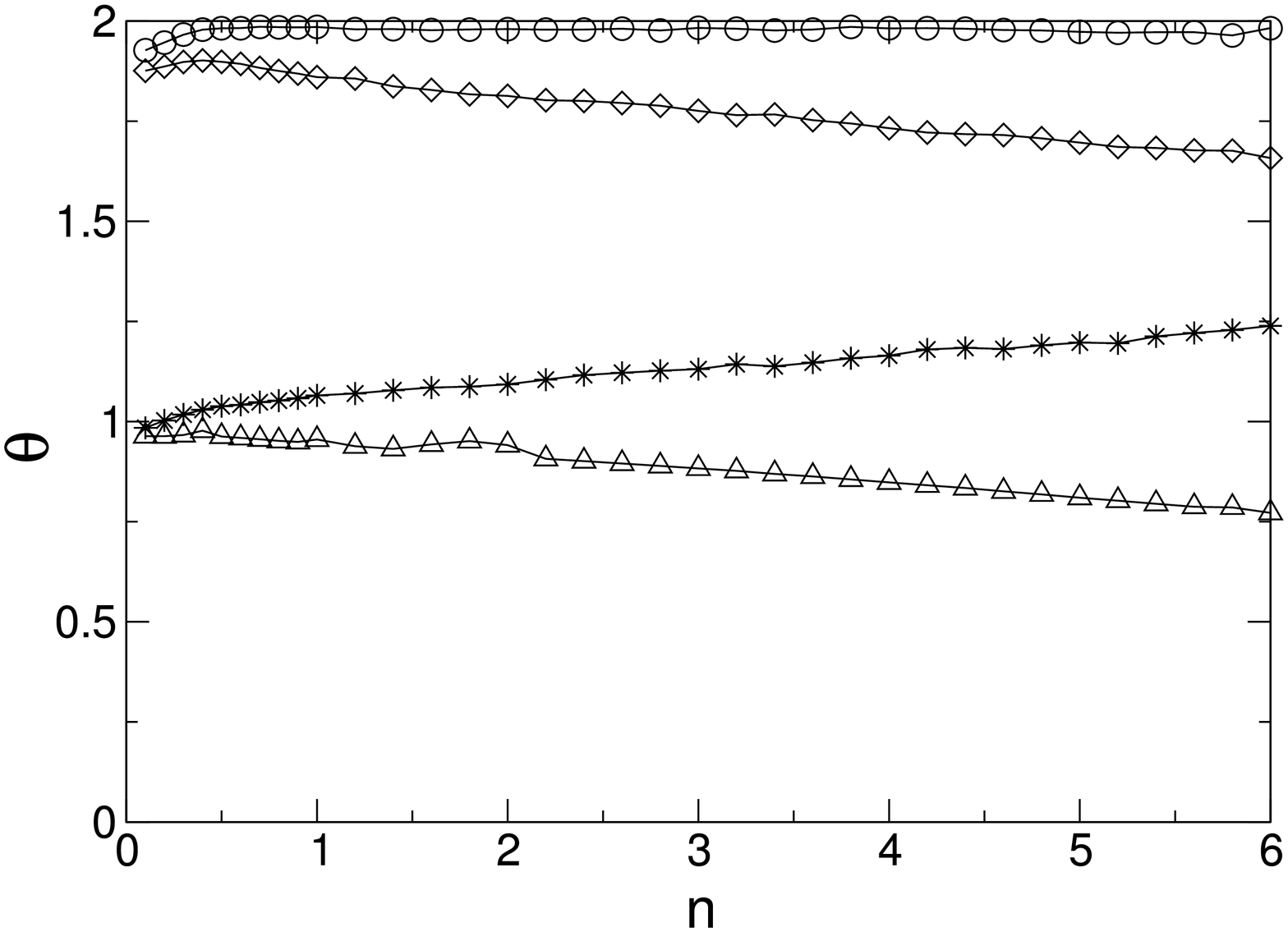} &
 \epsfysize=46mm \epsffile{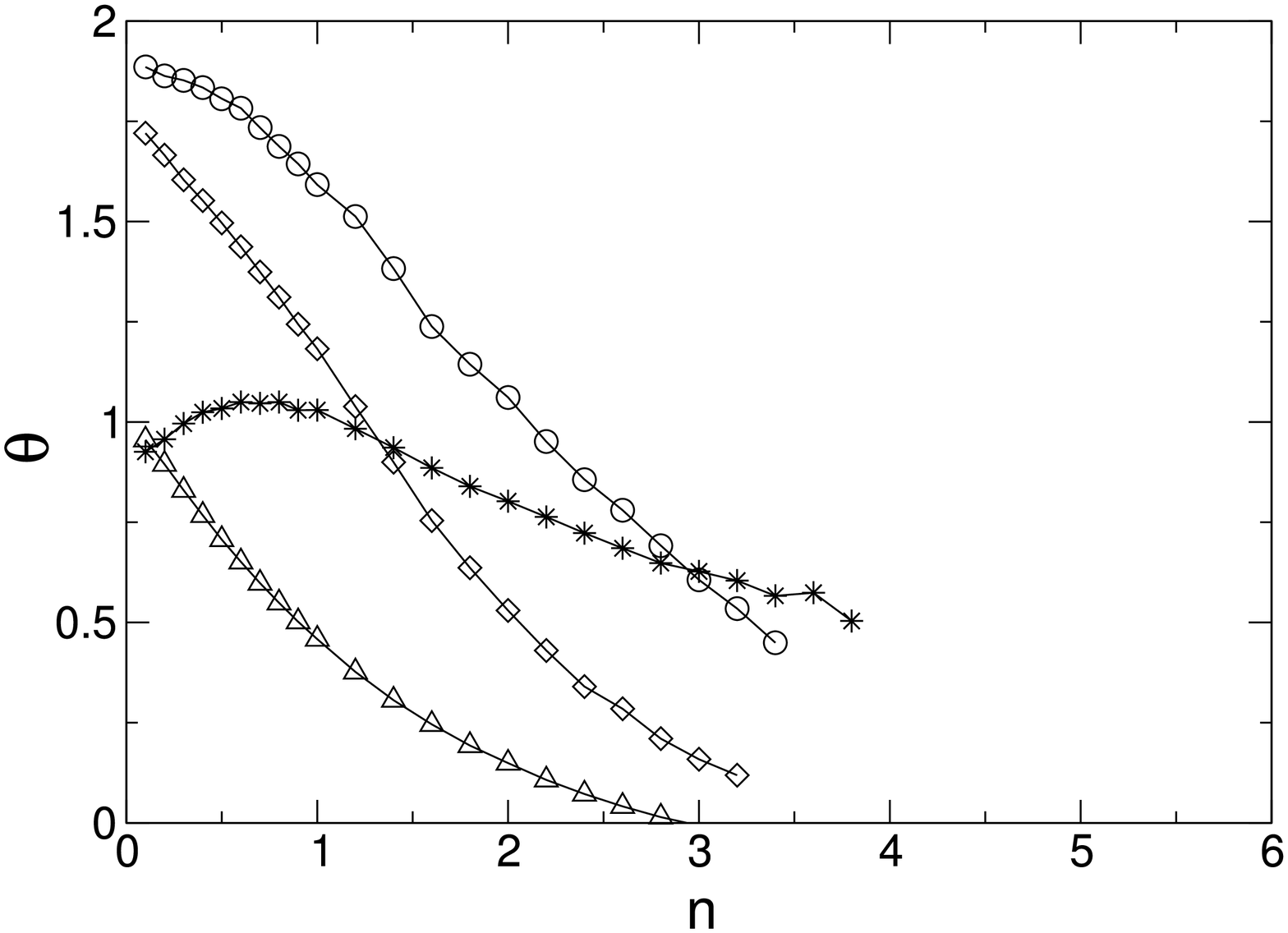} \\
\end{tabular}
}
\caption{ The exponents  $\theta$ as a
function of $n$ which give the scaling of the granular temperature $T$
($\Diamond$), collision frequency $N_c$ ($\ast$), mean impulsions $\Delta
I$ ($\bigtriangleup$) and pressure $P$ ($\circ$). All these quantities are
proportional to $V^{\theta(n)}$.  Without gravity: {\bf (a)} for $r=0.95$
and {\bf (b)} for $r=r(v)$. With gravity: {\bf (c)} for $r=0.95$ and {\bf
(d)} for $r=r(v)$.  The exponents are obtained by fixing $n$ and
performing eleven simulations, varying $f$ from $10$ Hz to $30$ Hz.
Then $\log X$ (where $X$ is the quantity being considered) is plotted
against $\log V$.  The resulting curve is always nearly a straight
line (except for $n>3$ in (d) -- see text),  and the exponent is 
calculated from a least squares fit.   
The pressure scaling point ($\bullet$) on (b)
is from experiment \cite{Falcon:99} performed in low-gravity. See Fig.~\ref{McNFal:fig:snapshots}a (resp. Fig.~\ref{McNFal:fig:snapshots}b) for typical snapshots corresponding to $n=3$, $g\ne0$ and $r=r(v)$ (resp. $r=0.95$)}
\label{McNFal:fig:thetap}
\end{figure*}

Many authors have postulated that the pressure on the upper wall $P$ (or
granular temperature $T$) is related to the piston velocity $V$
through $P,T \propto V^\theta$.  However, it is not clear what the
correct ``scaling exponent'' $\theta$ should be.  This question
 has been
addressed several times in the past, without a clear resolution of the
question \cite{McNamara:98,Huntley:98,Kumaran:98,Lee:95}. For example, kinetic
theory \cite{Warr:95,Kumaran:98} and hydrodynamic models \cite{Lee:95} predict
$T \propto V^{2}$ whereas numerical simulations \cite{Luding:94a,Luding:94b}
or experiments \cite{Warr:95,Yang:00,Luding:94a} give $T \propto V^{\theta}$,
with $1 \leq \theta \leq 2$. These studies were done at single values of
$n$.  In this section, we show that it is very important to
explicitly consider the dependence of the scaling exponents on $n$. 
We also consider the effect of gravity and a variable coefficient
of restitution.
Doing so enables us to explain and unify all previous works.

At the upper wall, we measured numerically the collision frequency, 
$N_c$ and the mean impulsion per
collision, $\Delta I$ for various
frequencies of the vibrating wall and numbers of particles in the box, with
$r=r(v)$ or with $r=0.95$, in the presence or absence of gravity.
The time averaged pressure on the upper wall
can be calculated from these quantities
using  
\begin{equation}
P=N_c \Delta I / L \ {\rm .}
\label{McNFal:eq:defP}
\end{equation}
(By conservation of momentum, the time averaged pressure on the
lower wall is just $P$ plus the weight of the particles $Nmg/L$.)
The total kinetic energy of the system is also measured to have access 
to the granular temperature, $T$. $N_c$, $\Delta I$, $P$,
and $T$ are all found
to fit with power laws in $V^{\theta}$ for our range of piston velocities.
Fig.\ \ref{McNFal:fig:thetap} shows $\theta$ exponents of $N_c$, 
$\Delta I$, $P$,
and  $T$ as a function of $n$. When $g=0$ and $r$ is constant (see Fig.\
\ref{McNFal:fig:thetap}a), we have $P\sim V^2$, $\Delta I \sim V$ 
and $N_c \sim V$
for all $n$. We call these relations the classical kinetic theory scaling.
This scaling can be established by simple dimensional analysis when the
vibration velocity $V$ provides the only time scale in the system. This is
the case for $g=0$ and $r$ independent of velocity. However, in the
experiments, two additional time scales are provided, one by gravity and
another by velocity dependent restitution coefficient. Numerical simulations
can separate the effects of these two new time scales on the scaling
exponents $\theta$. This is done in Fig.\ \ref{McNFal:fig:thetap}b [where $g=0$
but $r=r(v)$] and Fig.\ \ref{McNFal:fig:thetap}c [where $r$ is constant but
$g\ne0$]. In both figures, all the exponents become functions of $n$.
However, the time scale linked to $r=r(v)$ leads to much more dramatic
departure from the classical scaling. After considering the two time scale
separated, let us consider the case corresponding to most experiments,
where both gravity and $r=r(v)$ are present (see Fig.\ \ref{McNFal:fig:thetap}d).
The similarity between this figure and Fig.\ \ref{McNFal:fig:thetap}b confirms
that the velocity dependent restitution coefficient has a more important
effect than the gravity. Furthermore, only the variation of the restitution
coefficient on the particle velocity explains the experiment performed in
low-gravity \cite{Falcon:99}. This experiment gives a $V^{3/2}$ pressure
scaling ($\bullet$-mark on Fig.\ \ref{McNFal:fig:thetap}b) for $n=1$ and a
motionless clustered state for $n>2$. Only the simulation with $r=r(v)$ can
reproduce these results (see Fig.\ \ref{McNFal:fig:thetap}b) whereas constant $r$
simulations leads to the classical scaling ($P \propto V^{2}$, see Fig.\
\ref{McNFal:fig:thetap}a) and only a gaseous state for all $n$ shown in the figure.

\begin{figure}[ht]
\centerline{
 \epsfysize=50mm \epsffile{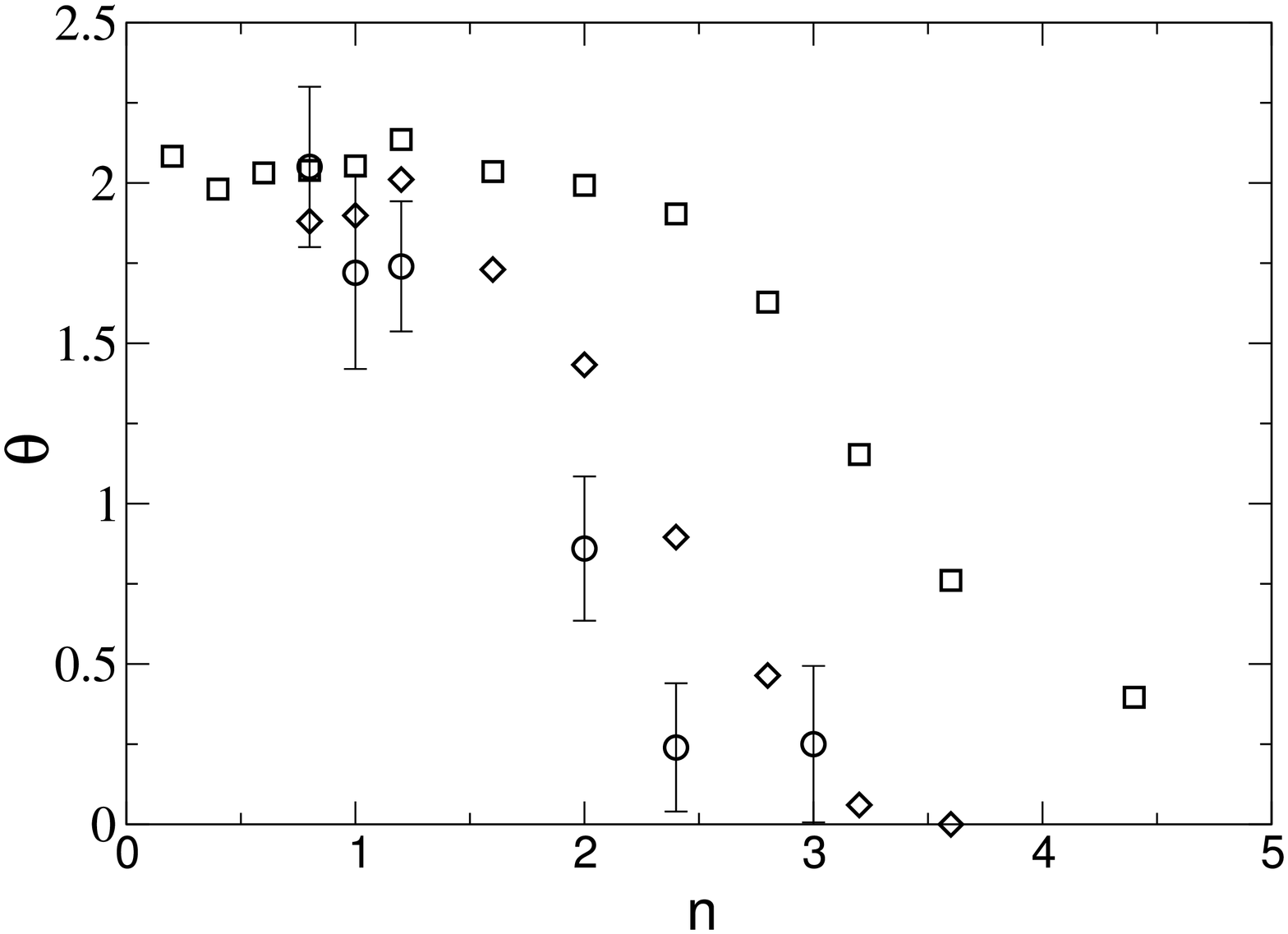}}
\caption[]{Experimental data performed under gravity from Ref. 
 \cite{Falcon:01}: The
exponents $\theta(n)$ of time averaged pressure ($\square$) (see
Fig.~\ref{McNFal:fig:freqscan}a), and kinetic energy extracted from
density profile ($\circ$) or volume expansion ($\lozenge$) measurements. 
These data should be compared with the simulations of 
Fig.~\ref{McNFal:fig:thetap}d.}
\label{McNFal:fig:eric}
\end{figure}

As shown in Fig\ \ref{McNFal:fig:thetap}, 
it is thus very important to explicitly consider the dependence of $\theta$
on $n$.  In all cases, except the unrealistic case of Fig.\
\ref{McNFal:fig:thetap}a, $\theta$ depends on $n$.  To our
knowledge, the only experiment  \cite{Falcon:01} to systematically
investigate this effect shows that  $T \propto V^{\theta(n)}$, with $\theta$
continuously varying from $\theta = 2$ when $n \rightarrow 0$, as expected
from kinetic theory, to $\theta \simeq 0$ for large $n$ due to the
clustering instability. These experiments \cite{Falcon:01} performed under
gravity (shown in Fig.\ \ref{McNFal:fig:eric}) are well reproduced
by the simulations of Fig.\ \ref{McNFal:fig:thetap}d. In both cases, the observed
pressure and granular temperature scaling exponents strongly decrease with
$n$. 

We finish this section by noting two curious facts about
Fig.~\ref{McNFal:fig:thetap}.  First of all, in Fig.~\ref{McNFal:fig:thetap}b
[$g=0$ and $r=r(v)$], $\theta\approx 1$ for the pressure and temperature
when $n>2$.  This is the sign of new robust scaling regime where $P$ and $T \propto V^1$,
which will be the topic of a future paper.
Secondly, in Fig.~\ref{McNFal:fig:thetap}d [$g \neq 0$ and $r=r(v)$], the scaling exponents are not
shown for $n\ge3$, because the dependence of $P$, $T$, $N_c$
and $\Delta I$ on $V$ is no longer a simple power law.  (More
precisely, we do not plot a point on
Fig.~\ref{McNFal:fig:thetap} when
$\left| \log X_{\text{observed}}-\log X_{\text{fitted}}\right| \ge 0.25$
for any of the eleven simulations used to calculate the exponent --
see caption.)
The power law breaks down because there is a resonance
between the time of flight of the cluster under gravity and the 
vibration period. 

\subsection{The influence of other parameters}
\label{otherparameters}

In this section, we review the influence of the other simulation parameters 
(box size, particle rotations, gravity, and the vibration parameters), and show that 
it is not possible to reproduce the experimental curves in Fig.~\ref{McNFal:fig:freqscan}a
unless one sets $r=r(v)$ or $r=0.7$.


Performing simulations for $5$ mm $\le h-h_0\le 50$ mm show that
the shapes of the curves $P$ vs. $n$ in Fig.~\ref{McNFal:fig:freqscan}b and \ref{McNFal:fig:freqscan}c remain the same.
For $r=r(v)$ (Fig.~\ref{McNFal:fig:freqscan}b) increasing $h-h_0$ shifts the maximum
towards smaller values of $n$ and decreases in amplitude.
The only exception occurs when box height approaches the particle
diameter, i.e.~$h-h_0=5$ mm, where the maximum disappears.  Considering $r=0.95$ 
leads to similar conclusions.

To eliminate the possibility that experimental curve can be reproduced by taking
into account particle rotations, we performed simulations with $r=0.95$ and
various values of the tangential restitution coefficient $r_t$.  This parameter is defined
as the ratio between the tangential components of the pre- and post-collision 
relative velocities.  Perfectly smooth spheres correspond to $r_t=-1$.  When
$r_t=1$, the tangential relative velocity is reversed by the collision.  These two
values, $|r_t|=1$, correspond to energy conservation.  Energy is dissipated for
$-1 < r_t < 1$, $r_t=0$ corresponding to maximum energy dissipation.
When $|r_t|$ is close to one, the $P$ vs. $n$ curves are almost
unchanged.  When $r_t$ is close to $0$, the curves become nearly flat for $n>2$.



Throughout this paper, we have used
the piston vibration velocity $V$ to characterize the vibration. It is important to 
point out that $V$ is not
the only way to do this.  One could also use the
maximum piston acceleration $\Gamma$.  When $\Gamma$ is close to $g$,
it controls the behavior of the system, i.e., adjusting $A$ and $f$
while keeping $\Gamma$ constant does not change the system's behavior
much.  But in the simulations presented here, $\Gamma\gg g$, and the
system's behavior is controlled by $V$.  This can be checked by
multiplying the frequency by $10$ while dividing $A$ by $10$,
thus keeping $V$ the same (while $\Gamma$ increases by an order
of magnitude).  Doing so changes the pressure only by about $20\%$.
Therefore, $V$ is the correct parameter to describe the vibration for the
simulations considered here.

\section{Conclusions}
\label{conclusions}

In this paper, we brought simulations of a strongly vibrated granular medium as close as possible to the experiments.  We show that the use of a velocity dependent coefficient of restitution reproduce results that agree with experiments.  It is especially important to take into account plastic deformations that cause the restitution coefficient to decrease rapidly with increasing impact velocity.  Indeed, the restitution coefficient for strongly vibrated steel spheres is very far from the constant values of $r=0.95$ or $r=0.9$ that are often cited in simulations as typical for steel spheres.  Changing the box size, the gravitational acceleration and including particle rotation do not modify this conclusion.
We also noted that
it is very important to take into account the number of particle layers $n$.  The dependence of the pressure $P$ on the piston velocity $V$ changes with $n$.  It
is not accurate to speak of ``a'' scaling exponent for the pressure in terms of $V$: this exponent depends continuously on $n$, and does not exist at high density ($n>3$) under gravity, due to the clustering instability.

\begin{acknowledgments}
We thank St\'ephan Fauve for fruitful discussions. The authors gratefully acknowledge the hospitality of the ENS-Lyon physics department which made this collaboration possible.\\
\end{acknowledgments}

\bibliographystyle{prsty}

\end{document}